\documentclass{article}
\usepackage{spconf,amsmath,graphicx}

\usepackage{enumitem}
\usepackage{multirow}
\usepackage{cite}
\usepackage{mathbbol}
\setlist{nosep, leftmargin=14pt}

\usepackage{mwe}

\usepackage{xcolor}

\title{GLFC: Unified Global-Local Feature and Contrast Learning with Mamba-Enhanced UNet for Synthetic CT Generation from CBCT}
\name{Xianhao Zhou$^1$, Jianghao Wu$^1$, Huangxuan Zhao$^2$, Lei Chen$^2$, Shaoting~Zhang$^{1,3}$, Guotai Wang$^{1,3}$\thanks{Corresponding author: Guotai Wang (guotai.wang@uestc.edu.cn).}}
\address{$^1$ School of Mechanical and Electrical Engineering,\\ 
University of Electronic Science and Technology of China, Chengdu, China \\
$^2$ Department of Radiology, Union Hospital, Tongji Medical College, \\ 
Huazhong University of Science and Technology, Wuhan, China \\
$^3$ Shanghai Artificial Intelligence Laboratory, Shanghai, China \\
}

\begin{document}

\maketitle

\begin{abstract}
Generating synthetic Computed Tomography (CT) images from 
Cone Beam Computed Tomography (CBCT) is desirable for improving the image quality  of CBCT. Existing synthetic CT (sCT) generation methods using Convolutional Neural Networks (CNN) and Transformers often face difficulties in effectively capturing both global and local features and contrasts for high-quality sCT generation. 
In this work, we propose a Global-Local Feature and Contrast learning (GLFC) framework for sCT generation. First, a Mamba-Enhanced UNet (MEUNet) is introduced by integrating Mamba blocks into the skip connections of a high-resolution UNet for effective global and local feature learning. Second, we propose a Multiple Contrast Loss (MCL) that calculates synthetic loss at different intensity windows to improve  quality for both soft tissues and bone regions. Experiments on the SynthRAD2023 dataset demonstrate that GLFC improved the SSIM of sCT from 77.91\% to 91.50\% compared with the original CBCT, and significantly outperformed several existing methods for sCT generation. The code is available at https://github.com/HiLab-git/GLFC.

\end{abstract}

\begin{keywords}
Cone Beam CT (CBCT), Image translation, UNet, Mamba, Multiple contrast loss.
\end{keywords}

\section{Introduction}
\label{sec:intro}

Cone Beam Computed Tomography (CBCT) is widely employed in various clinical applications 
due to its low implementation costs, reduced radiation exposure, and rapid imaging capabilities~\cite{cbct_application_review}. However, CBCT images are often hampered by significant artifacts and inaccurate Hounsfield Unit (HU) values compared to conventional Computed Tomography (CT), which limits its effectiveness in critical applications such as precise dose calculation for radiotherapy and tumor assessment~\cite{cbct_application_review}. 
Generating synthetic CT (sCT) from CBCT is a potential solution for this problem, and it allows clinicians to harness the cost-effectiveness of CBCT while achieving the superior image quality associated with CT~\cite{sct_review,cbct2sct_review}.

In recent years, deep learning methods have been widely used for sCT generation from CBCT. However, many existing approaches encounter challenges in effectively capturing both global and local features and contrasts due to inherent limitations in their network architectures and loss functions~\cite{sct_review,cbct2sct_review,sct_ISBI}. For network architecture, Convolutional Neural Networks (CNN) such as UNet~\cite{Unet} have a limited receptive field in convolutional layers for global feature learning~\cite{Unet}. Vision transformers are better at modeling global features due to the self-attention mechanism, but are impeded by quadratic computational complexity and limited ability to recovery local details~\cite{ViT,SwinT}. Recently, Mamba, a sequence model based on State Space Models (SSM), has gained attention for its ability to model long sequences with a global perspective while maintaining linear computational complexity~\cite{mamba}. This has led to applying Mamba to vision tasks~\cite{Vim,Vmamba} such as image segmentation and generation~\cite{mamba-Unet,I2I-mamba} by replacing convolutional layers of UNet variants with vision Mamba blocks. Despite their improved global feature learning ability, they sacrifice the convolution's inherent advantage in capturing local features. In addition, these methods often use multiple down-sampling layers with increased channel numbers,  which reduces the image resolution for fine-grained image generation with increased model complexity. 

In terms of loss functions, existing methods typically normalize the full HU range (i.e., around -1000 to 3000 of CT) to [0, 1] or [-1,1] for loss calculation at a global window
~\cite{cbct2ct_Unet,cbct2ct_cGAN,cbct2ct_cycleGAN}. Though this helps to obtain satisfactory outputs at a global view, it may lead to poor performance for a specific HU range relevant to some important tissues. 
For instance, soft tissues in the brain such as gray matter and white matter have a narrow HU range around [-250, 250], and bones like the skull occupy HU values in the range around [300, 2000]. Using the global window with full HU range for loss calculation will make the network pay insufficient attention to such tissues, leading to limited synthesis quality for them. 

To deal with these problems, we propose a novel Global-Local Feature and Contrast (GLFC) learning framework for generating sCT images from CBCT. 
First, to better leverage both global and local features for synthesis, we propose a Mamba-Enhanced UNet (MEUNet) that integrates Mamba blocks into skip connections of a UNet with only two down-sampling layers. The convolutions in the encoder and decoder with  reduced down-sampling layers help to keep high-resolution for synthesizing local details, and Mamba in the skip connection helps to capture long-range dependency with improved awareness of global semantics. 
Second, we introduce a Multiple Contrast Loss (MCL) that combines global and local intensity windows for loss calculation, which enhances the accuracy of critical structures like soft tissues and bones with local HU ranges while maintaining strong global consistency.
Experiments on the SynthRad2023 dataset demonstrated that both MEUNet and MCL effectively improved the quality of synthetic CT images in terms of SSIM and PSNR, and our GLFC framework outperformed several state-of-the-art deep learning methods for sCT generation.

\section{METHODS}
As shown in Fig.~\ref{fig0}, our GLFC framework uses a  Mamba-Enhanced UNet (MEUNet) trained with a Multiple Contrast Loss (MCL) for sCT generation, where MEUNet 
combines the advantage of vision Mamba and convolutional blocks in a high-resolution UNet, and MCL calculates loss values with a global intensity window and two local intensity windows to improve the quality of synthesized soft tissues and bones. 

\subsection{MEUNet for Global-Local Feature Extraction}
To improve global feature learning ability, CNNs~\cite{Unet} use multiple down-sampling layers to enlarge the receptive field. However, it  not only reduces the resolution of feature maps and limits the performance on generating local details, but also improves the model complexity by exponentially increasing channel numbers after each down-sampling. Motivated by Mamba's superiority on learning global features from long sequences, we keep high-resolution of feature maps in MEUNet and use Mamba for long-range dependency learning rather than applying too many down-sampling layers.  

\begin{figure}
\begin{minipage}{1.0\linewidth}
  \centering
  \centerline{\includegraphics[width=8.8cm]{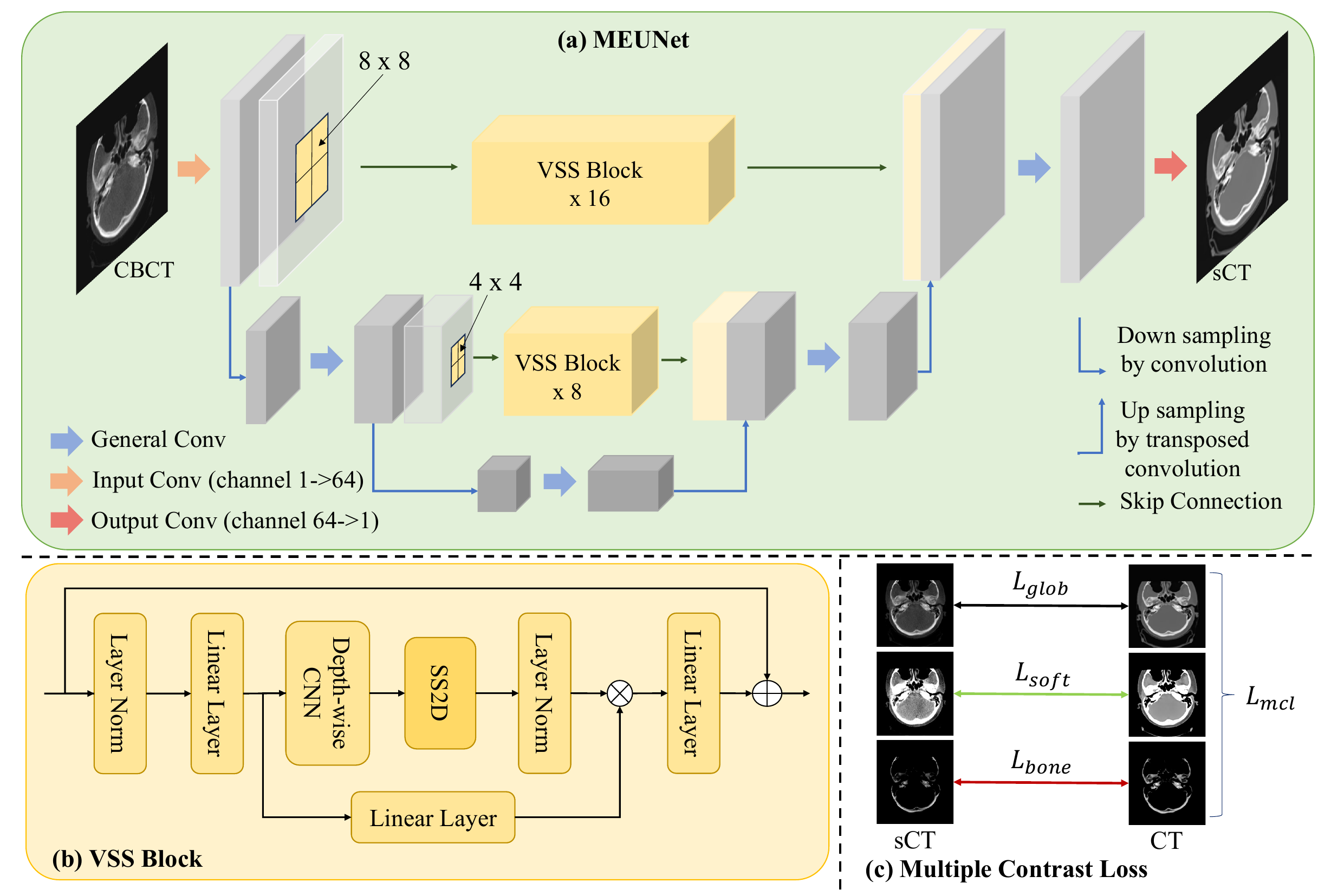}}
\caption{Overview of our Global-Local Feature and Contrast  (GLFC) learning framework for synthetic CT generation. 
}
\label{fig0}
\end{minipage}
\end{figure}

As shown in Fig.~\ref{fig0}(a),  MEUNet only has two down-sampling layers in the encoder, and before each down-sampling, a convolutional block with two convolutional layers is used to extract local features.    
To obtain high-quality synthesis of local details in the output, we also use convolutional layers in the decoder. Importantly, to improve global feature representation, we incorporate Visual State Space (VSS)~\cite{Vmamba} blocks of Mamba into the skip connections between the encoder and decoder. In VSS, each patch is treated as a token, and to keep the same number of tokens at the two skip connections, we use an adaptive patching strategy. Specifically, let $N_i \times N_i$ denote the spatial dimension of the feature map at the $i$-th resolution level ($i$ = 0, 1), we use $L$ to denote the predefined number of tokens, the patch size for the $i$-th resolution level is denoted as $M_i \times M_i$, where  $M_i = \sqrt{N_i^2 / L}$. 
\label{sec:meth}


In each skip connection, the $L$ patches are then processed by VSS blocks, as shown in Fig.~\ref{fig0}(b). The core of VSS is a SS2D module that learns the relationship among a sequence of tokens by a discrete State Space Model (SSM)~\cite{Vmamba}: 

\begin{equation} \begin{gathered}
h_t=\bar{A} h_{t-1}+\bar{B} x_t \\ 
y_t=C h_t \\ \bar{A} =e^{\Delta A} \\
\bar{B} = (e^{\Delta A}-I)A^{-1}  B 
 \end{gathered} \end{equation}
where $x_t$,  $y_t$ and $h_t$ are the input, output and hidden state for the $t$-th token, respectively. Let $D$ denote the hidden state dimension, the matrices $A \in \mathbb{R}^{D\times D}$, $B\in \mathbb{R}^{D\times1}$, $C \in \mathbb{R}^{1\times D}$ and the scalar $\Delta$ are either learnable parameters or values computed based on real-time 
input $x$ and learnable parameters. Note that the input of the first skip connection is a relatively low-level feature, we use more VSS blocks there than the second skip connection, i.e., 16 and 8 VSS blocks are used in the two skip connections, respectively. 

\subsection{Multiple Contrast Loss (MCL)}
In this work, we normalize the image intensity to [-1.0, 1.0] for network prediction $P$ and ground truth CT $Y$ based on the full HU range.  Typical supervised image translation methods directly use $P$ and $Y$ to calculate a global loss~\cite{cbct2sct_review}, e.g., $L_{glob} = ||Y-P||_1$. 
To better capture details for tissues with narrower HU ranges, we introduce MCL that also calculates the loss at multiple local intensity windows. 

Let $\mathbf{w}_n = [I^n_0, I^n_1]$ denote the $n$-th local intensity window with the min and max value being  $I^n_0$ and  $I^n_1$, respectively. $P$ is first normalized by  $\mathbf{w}_n$ as  $P'_n = 2(P-I^n_0) / (I^n_1 - I^n_0) -1$, and then clipped to the range of [-1.0,1.0], obtaining $\hat{P_n} = Clip(P'_n)$. Correspondingly, $Y$ is also normalized by $\mathbf{w}_n$ and clipped to [-1.0,1.0], and the result is denoted as $\hat{Y}_n$. The local window loss based on   $\mathbf{w}_n$ is  defined as $L_n = ||\hat{Y}_n - \hat{P}_n||_1$. 

For sCT generation, we use two local intensity windows: The first one is for soft tissues $\mathbf{w}_{soft}$ = [-0.615, -0.368], which corresponds to the HU range of [-250, 250]. The second one is for bones $\mathbf{w}_{bone}$ = [-0.368, 1.0], which corresponds to the HU range of [250, 3000]. The local window losses corresponding to  $\mathbf{w}_{soft}$ and $\mathbf{w}_{bone}$ are denoted as $L_{soft}$ and $L_{bone}$, respectively. The overall MCL loss is defined as: 

\begin{equation} 
L_{mcl} = L_{glob} + L_{soft} + L_{bone} 
\end{equation}
where $ L_{glob} $ is a loss calculated in a global contrast based on the full HU range of [-1024, 3000], and $L_{soft}$ and $L_{bone}$ encourages better contrasts for soft tissues and bones based on local HU ranges, respectively. 

\section{EXPERIMENTS AND RESULTS}
\label{sec:exp}
\subsection{Data and Implementation}

The public SynthRAD2023 Grand Challenge dataset was used for experiments, and it was  designed for generating synthetic CT images from CBCT for radiotherapy~\cite{SynthRAD2023}. The dataset consists of images from 180 patients, with each patient having a  pair of 3D head and neck CBCT and CT images that have been registered. The resolution is 1$\times$1$\times$1 $mm^3$, with a median image dimension of 256$\times$256$\times$200. 
We randomly split the dataset into 140, 20 and 20 pairs for training, validation, and testing respectively. From the 140 training cases, we extracted 28,631 pairs of 2D slices to train our 2D MEUNet implemented by PyTorch. Each slice was resized to 256$\times$256, and normalized to [-1.0,1.0]. The feature map channel number at the three resolution levels  was 64, 128, and 256, respectively. The length  of  token was $L$=1024 for the VSS blocks. All experiments were conducted on an NVIDIA 2080Ti GPU with a batch size of 4. The model was trained using the Adam optimizer with a learning rate of 0.02, running for 100 epochs.

For quantitative evaluation of sCT, we employed Structural Similarity Index (SSIM) and Peak Signal-to-Noise Ratio (PSNR) compared with real CT images. To focus more on the synthesis quality for human tissues, we ignored the air background, and calculated SSIM and PSNR only within the human region. In addition, we calculated SSIM and PSNR for soft tissue  and bone regions respectively to analyze the synthesis quality for different structures of interest.  

\subsection{Ablation Study of MEUNet}

\begin{table}\caption{
Ablation study of MEUNet. D$N$ refers to UNet with $N$ down-sampling layers. MEUNet(v1) and MEUNet(v2) mean using VSS blocks only in the first and second skip connection, respectively. MEUNet(f) uses a fixed patch size of 8$\times$8 for VSS, while our MEUNet uses adaptive patch size.
}\label{table1}
\scalebox{0.88}{
\begin{tabular}{l|ccc|ccc}
\hline
\multirow{2}{*}{Method}  &\multicolumn{3}{c|}{SSIM (\%)} &\multicolumn{3}{c}{PSNR} \\ 
\cline{2-7}
& full &ST &bone & full &ST &bone \\
\hline 
CBCT  & 77.91 & 39.27 &71.27 &20.09 &2.78 &15.61\\ 
\hline
UNet(D4) &90.01 & 51.30 &82.20 &28.84 &12.38 &21.81  \\
UNet(D3) &88.60 & 44.83 & 81.52 &28.07 &11.20 &21.58 \\
UNet(D2) &78.99 & 31.11 & 77.86 &24.90 &7.54 & 20.44 \\
MEUNet(v1) &90.17 &52.71 &82.86 & 28.99 & 12.64 & 21.77 \\
MEUNet(v2) &90.43 &53.18 &82.64 & 29.11 & 12.80 & 21.88 \\
MEUNet(f) &90.35 &52.79 &82.50 & 29.06 & 12.69 & 21.88 \\
MEUNet &\textbf{90.47} &\textbf{53.18} &\textbf{82.68} &\textbf{29.17} &\textbf{12.82} &\textbf{21.93} \\
\hline
\end{tabular}
}
\end{table}
To demonstrate the effectiveness of incorporating VSS blocks into the skip connections of a high-resolution UNet, we conducted an ablation study for the network structure using a typical global intensity window loss $L_{glob}$. We firstly compared different variants of UNet:   UNet(D2), UNet(D3) and UNet(D4)  that use 2, 3 and 4 down-sampling layers for UNet, respectively. 
The results in Table~\ref{table1} show that all these variants improved image quality from the original CBCT, and using a fewer number of down-sampling generally performed worse than a larger number of down-sampling for UNet, mainly due to that the later has a better global feature learning ability. However, when adding VSS blocks to the skip connections of UNet(D2), i.e., our MEUNet, the SSIM values improved from 78.99\% to 90.47\% for the entire human region, and from 31.11\% to 53.18\% for soft tissues. In addition, the performance of MEUNet is even better than UNet(D4) in terms of overall SSIM (90.47\% vs 90.01\%) and PSNR (29.17 vs 28.84). The comparison demonstrates that the VSS blocks in MEUNet can remove the need of using more down-sampling layers for global feature learning, and it also leads to better image quality than the typical UNet.

Moreover, we evaluated the impact of using VSS blocks in each of the two skip connections. Let MEUNet (v1) and  MEUNet (v2) denote using VSS blocks only in the fist and second skip connection, respectively.  The results in Table~\ref{table1} show that both variants already outperformed the original UNet in terms of both SSIM and PSNR, but they performed worse than our MEUNet that uses VSS blocks at two skip connections. 
Furthermore, by replacing our adaptive patching strategy for VSS with a fixed patch size, which is denoted as MEUNet(f), a decrease of SSIM and PSNR can be observed in Table~\ref{table1}, showing the superiority of adaptive patch size for MEUNet. 

\begin{figure*}
    \centering
    \includegraphics[width=1\linewidth]{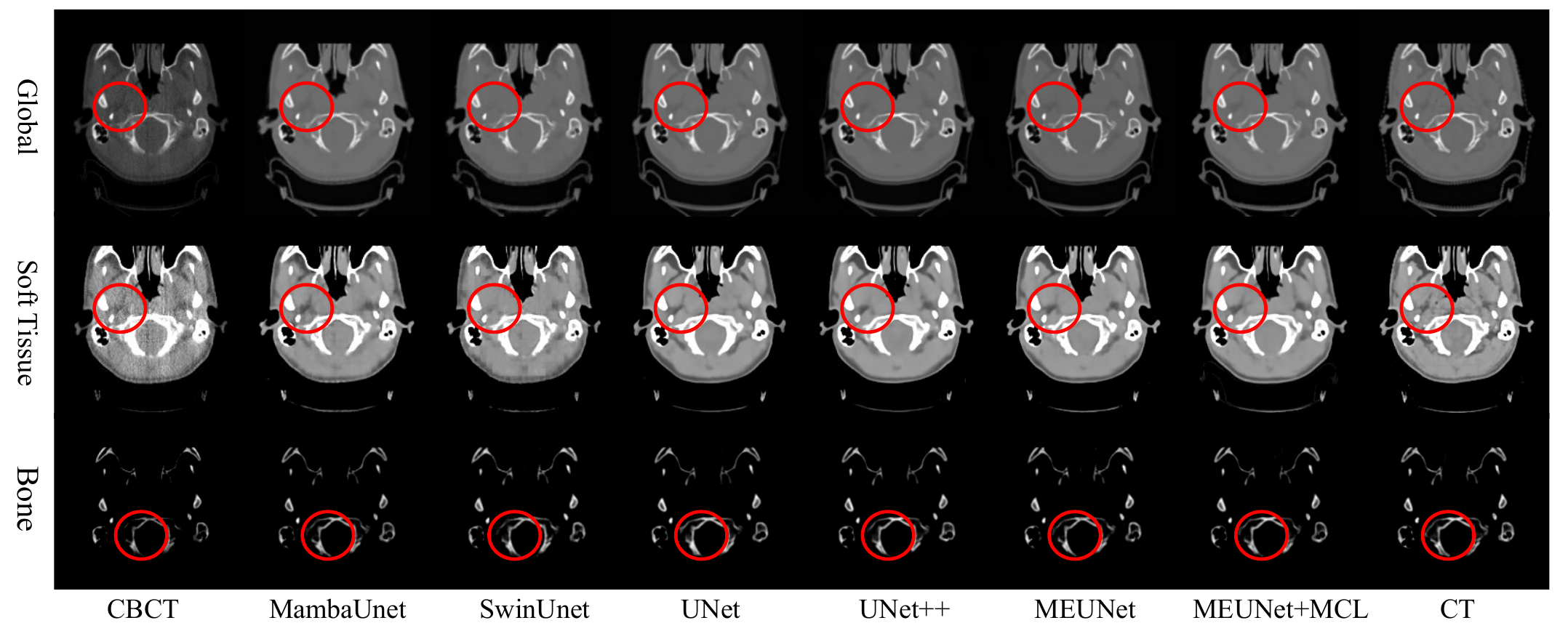}
    \caption{Visual comparison of sCT obtained by different methods. Images in the three rows are visualized with a global intensity window, soft tissue window and bone window, respectively. Local differences are highlighted by circles.  
    }
    \label{fig:enter-label}
\end{figure*}
\subsection{Comparison with Existing Methods}
We compared our method with four existing image translation networks: UNet~\cite{Unet}, UNet++~\cite{Unet++}, SwinUnet~\cite{SwinUnet}, and MambaUnet~\cite{mamba-Unet}. All the compared methods used the same loss function $L_{glob}$, and were trained with the same hyper-parameters as our method. 

Table~\ref{table2} shows a quantitative evaluation of these methods in the entire human region (full), soft tissue (ST) and bone, respectively. All the compared methods obtained relatively high SSIM and PSNR for the entire human region, but these metrics are much lower for the ST region, indicating the challenges in synthesizing soft tissues in sCT. Among the existing methods, UNet++~\cite{Unet++} obtained the best overall SSIM and PSNR value of 90.33\% and 28.97, respectively. In contrast, our MEUNet obtained a corresponding SSIM and PSNR value of 90.47\% and 29.17, respectively. Though the improvement of SSIM calculated in the entire human region is relatively slight, our MEUNet has a more obvious superiority over  UNet++~\cite{Unet++}  in dealing with the soft tissue region (53.18\% vs 51.79\% in terms of SSIM). In addition,  the size of MEUNet is only 1/9 that of UNet++ (5.57MB vs 47.1MB).

We then compared our $L_{mcl}$ with $L_{glob}$ under different backbone networks.  For MEUNet, replacing  $L_{glob}$ by $L_{mcl}$ improved the overall SSIM from 90.47\% to 91.50\%. In addition, $L_{glob}$ improved the PSNR values for both the entire human region and local target regions of soft tissues and bones. Table~\ref{table2}  shows that $L_{mcl}$ also improves the performance of UNet in terms of SSIM and PSNR calculated in different regions, demonstrating the generalizability of our MCL loss across different network architectures. 

\begin{table}\caption{
Quantitative comparison of different methods for sCT generation. ST: Soft tissues. 
* denotes a significant improvement (p-value $<$ 0.05) from the best existing method using a paired Student’s t-test.
}\label{table2}
\scalebox{0.69}{
\begin{tabular}{l|l|lll|lll}
\hline
\multirow{2}{*}{Network} &\multirow{2}{*}{Loss} &\multicolumn{3}{c}{SSIM (\%)} &\multicolumn{3}{|c}{PSNR} \\ 
\cline{3-8}
& & full &ST &bone & full &ST &bone \\
\hline 
\multicolumn{2}{c|}{CBCT} &77.91 & 39.97 &71.27 &20.09 &2.78 &15.61\\ 
\hline
MambaUnet~\cite{mamba-Unet} &$L_{glob}$ &89.41 &49.88 &80.91 &28.52 &12.07 &21.47  \\
SwinUnet~\cite{SwinUnet}  &$L_{glob}$ &88.83 &47.68 &79.80 &28.16 &11.71 &21.14 \\
UNet~\cite{Unet}  &$L_{glob}$ &90.00 & 51.20 &81.97 &28.80 &12.38 & 21.69 \\
UNet++~\cite{Unet++} &$L_{glob}$ &90.33 & 51.79 & 82.48 & 28.97 & 12.48 & 21.91 \\
\hline
MEUNet &$L_{glob}$ &90.47 &53.18 &82.68 &29.17 &12.82 &21.93 \\
UNet &$L_{mcl}$ &90.72 &52.37 &82.44 &29.27 &13.10 &21.82   \\
MEUNet &$L_{mcl}$ &\textbf{91.50*} &\textbf{56.83*} &\textbf{83.00*} &\textbf{29.84*} &\textbf{13.91*} &\textbf{22.06*}  \\
\hline
\end{tabular}
}
\end{table}

Fig.~\ref{fig:enter-label} shows a visual comparison of sCT obtained by  different methods. 
It demonstrates that the images produced by GLFC closely match the real CT images under a global intensity window. As the global window leads to low contrast of images and makes it hard to observe the differences in the compared methods, we also show the results with a soft tissue window and a bone window.  
In the soft tissue window, it can be observed that our method obtains a better soft tissue quality than existing methods such as SwinUNet and UNet++.

\section{CONCLUSION}
In conclusion, this work introduces a novel framework based on Global-Local Feature and Contrast (GLFC)  learning for synthesizing CT images from CBCT. It comprises a Mamba-Enhanced UNet (MEUNet) and a Multiple Contrast Loss (MCL). MEUNet effectively integrates Mamba's long-sequence modeling capabilities with UNet, enabling the capture of both global and local features. The MCL simultaneously considers a global contrast and two local contrasts  that highlight key regions such as soft tissues and bones. Experimental results demonstrate that our approach achieved state-of-the-art performance on the SynthRAD2023 dataset. This advancement has the potential to improve CBCT's diagnostic accuracy and treatment planning in radiotherapy, and it is of interest to investigate the performance of our method on other medical image translation tasks in future works. 
\label{sec:dis}


\section{compliance with ethical standards}
This research study was conducted retrospectively using human subject data made available in open access. Ethical approval was not required as confirmed by the license attached with the open-access data.

\label{sec:ack}

\bibliographystyle{IEEEbib}
\bibliography{strings,refs}

\begin{thebibliography}{10}

\bibitem{cbct_application_review}
Keith Horner, Lucy O'Malley, Kathryn Taylor, and Anne-Marie Glenny,
\newblock ``Guidelines for clinical use of {CBCT}: a review,''
\newblock {\em Dentomaxillofacial radiology}, vol. 44, no. 1, pp. 20140225, 2015.

\bibitem{sct_review}
Maria~Francesca Spadea, Matteo Maspero, Paolo Zaffino, and Joao Seco,
\newblock ``Deep learning based synthetic-{CT} generation in radiotherapy and {PET}: a review,''
\newblock {\em Medical physics}, vol. 48, no. 11, pp. 6537--6566, 2021.

\bibitem{cbct2sct_review}
Branimir Rusanov, Ghulam~Mubashar Hassan, Mark Reynolds, Mahsheed Sabet, Jake Kendrick, Pejman Rowshanfarzad, and Martin Ebert,
\newblock ``Deep learning methods for enhancing cone-beam {CT} image quality toward adaptive radiation therapy: A systematic review,''
\newblock {\em Medical Physics}, vol. 49, no. 9, pp. 6019--6054, 2022.

\bibitem{sct_ISBI}
Axel Largent, Jean-Claude Nunes, Hervé Saint-Jalmes, John Baxter, Peter Greer, Jason Dowling, Renaud~de Crevoisier, and Oscar Acosta,
\newblock ``Pseudo-{CT} generation for mri-only radiotherapy: Comparative study between a generative adversarial network, a {U-Net} network, a patch-based, and an atlas based methods,''
\newblock in {\em ISBI}, 2019, pp. 1109--1113.

\bibitem{Unet}
Olaf Ronneberger, Philipp Fischer, and Thomas Brox,
\newblock ``{U-Net}: Convolutional networks for biomedical image segmentation,''
\newblock in {\em MICCAI}, 2015, pp. 234--241.

\bibitem{ViT}
Alexey Dosovitskiy,
\newblock ``An image is worth 16x16 words: Transformers for image recognition at scale,''
\newblock {\em arXiv preprint arXiv:2010.11929}, 2020.

\bibitem{SwinT}
Ze~Liu, Yutong Lin, Yue Cao, Han Hu, Yixuan Wei, Zheng Zhang, Stephen Lin, and Baining Guo,
\newblock ``Swin transformer: Hierarchical vision transformer using shifted windows,''
\newblock in {\em ICCV}, 2021, pp. 10012--10022.

\bibitem{mamba}
Albert Gu and Tri Dao,
\newblock ``Mamba: Linear-time sequence modeling with selective state spaces,''
\newblock {\em arXiv preprint arXiv:2312.00752}, 2023.

\bibitem{Vim}
Lianghui Zhu, Bencheng Liao, Qian Zhang, Xinlong Wang, Wenyu Liu, and Xinggang Wang,
\newblock ``Vision mamba: Efficient visual representation learning with bidirectional state space model,''
\newblock {\em arXiv preprint arXiv:2401.09417}, 2024.

\bibitem{Vmamba}
Yue Liu, Yunjie Tian, Yuzhong Zhao, Hongtian Yu, Lingxi Xie, Yaowei Wang, Qixiang Ye, and Yunfan Liu,
\newblock ``Vmamba: Visual state space model 2024,''
\newblock {\em arXiv preprint arXiv:2401.10166}, 2024.

\bibitem{mamba-Unet}
Ziyang Wang, Jian-Qing Zheng, Yichi Zhang, Ge~Cui, and Lei Li,
\newblock ``Mamba-unet: {Unet}-like pure visual mamba for medical image segmentation,''
\newblock {\em arXiv preprint arXiv:2402.05079}, 2024.

\bibitem{I2I-mamba}
Omer~F Atli, Bilal Kabas, Fuat Arslan, Mahmut Yurt, Onat Dalmaz, and Tolga {\c{C}}ukur,
\newblock ``I2i-mamba: Multi-modal medical image synthesis via selective state space modeling,''
\newblock {\em arXiv preprint arXiv:2405.14022}, 2024.

\bibitem{cbct2ct_Unet}
Liyuan Chen, Xiao Liang, Chenyang Shen, Steve Jiang, and Jing Wang,
\newblock ``Synthetic {CT} generation from {CBCT} images via deep learning,''
\newblock {\em Medical physics}, vol. 47, no. 3, pp. 1115--1125, 2020.

\bibitem{cbct2ct_cGAN}
Jiwei Liu, Hui Yan, Hanlin Cheng, Jianfei Liu, Pengjian Sun, Boyi Wang, Ronghu Mao, Chi Du, and Shengquan Luo,
\newblock ``{CBCT}-based synthetic {CT} generation using generative adversarial networks with disentangled representation,''
\newblock {\em Quantitative Imaging in Medicine and Surgery}, vol. 11, no. 12, pp. 4820, 2021.

\bibitem{cbct2ct_cycleGAN}
Yingzi Liu, Yang Lei, Tonghe Wang, Yabo Fu, Xiangyang Tang, Walter~J Curran, Tian Liu, Pretesh Patel, and Xiaofeng Yang,
\newblock ``{CBCT}-based synthetic {CT} generation using deep-attention cyclegan for pancreatic adaptive radiotherapy,''
\newblock {\em Medical physics}, vol. 47, no. 6, pp. 2472--2483, 2020.

\bibitem{SynthRAD2023}
Adrian Thummerer, Erik van~der Bijl, Arthur Galapon~Jr, Joost~JC Verhoeff, Johannes~A Langendijk, Stefan Both, Cornelis (Nico)~AT van~den Berg, and Matteo Maspero,
\newblock ``Synthrad2023 grand challenge dataset: Generating synthetic {CT} for radiotherapy,''
\newblock {\em Medical physics}, vol. 50, no. 7, pp. 4664--4674, 2023.

\bibitem{Unet++}
Zongwei Zhou, Md~Mahfuzur Rahman~Siddiquee, Nima Tajbakhsh, and Jianming Liang,
\newblock ``Unet++: A nested u-net architecture for medical image segmentation,''
\newblock in {\em MICCAI DLMIA Workshop}, 2018, pp. 3--11.

\bibitem{SwinUnet}
Hu~Cao, Yueyue Wang, Joy Chen, Dongsheng Jiang, Xiaopeng Zhang, Qi~Tian, and Manning Wang,
\newblock ``Swin-unet: {Unet}-like pure transformer for medical image segmentation,''
\newblock in {\em ECCV}, 2022, pp. 205--218.

\end{thebibliography}

\end{document}